\documentclass[pra,twocolumn]{revtex4}
\usepackage{graphicx}
\usepackage{amsfonts,amsmath,amssymb,float}
\usepackage{bm,dsfont}
\usepackage{color}
\usepackage{bbm}
\usepackage{longtable}
\usepackage[dvips]{epsfig}
\usepackage{amsmath,amssymb,lscape,float}
\usepackage{hyperref}
\usepackage{listings}

\begin{document}
\title{Scalable $W$-type entanglement resource in neutral-atom arrays 
with Rydberg-dressed resonant dipole-dipole interaction}

\author{Vladimir M. Stojanovi\'c}
\affiliation{Institut f\"{u}r Angewandte Physik, Technical
University of Darmstadt, D-64289 Darmstadt, Germany}

\date{\today}

\begin{abstract}
While the Rydberg-blockade regime provides the natural setting for creating $W$-type entanglement with cold neutral atoms, 
it is demonstrated here that a scalable entanglement resource of this type can even be obtained under completely different
physical circumstances. To be more precise, a special instance of twisted $W$ states -- namely, $\pi$-twisted ones -- can 
be engineered in one-dimensional arrays of cold neutral atoms with Rydberg-dressed resonant dipole-dipole interaction. 
In particular, it is shown here that this is possible even when a (dressed) Rydberg excitation is coupled to the motional degrees 
of freedom of atoms in their respective, nearly-harmonic optical-dipole microtraps, which are quantized into dispersionless 
(zero-dimensional) bosons. For a specially chosen (``sweet-spot'') detuning of the off-resonant dressing lasers from the relevant 
internal atomic transitions, the desired $\pi$-twisted $W$ state of Rydberg-dressed qubits is the ground state of the effective 
excitation -- boson Hamiltonian of the system in a broad window of the relevant parameters. Being at the same time separated from 
the other eigenstates by a gap equal to the single-boson energy, this $W$ state can be prepared using a Rabi-type driving protocol. 
The corresponding preparation times are independent of the system size and several orders of magnitude shorter than the effective 
lifetimes of the relevant atomic states.
\end{abstract}

\maketitle
\section{Introduction}
The last decade has seen a surge of interest in ensembles of Rydberg atoms~\cite{GallagherBOOK}, both from the 
fundamental-physics~\cite{Weber+:17} and technological standpoints~\cite{Adams+:20}. Owing to the remarkable 
properties of their atomic constituents, these systems have gained prominence as analog simulators of many-body 
phenomena~\cite{Tamura+:20,Carollo+:20,Browaeys+Lahaye:20}. Large strides have simultaneously been made in the 
context of quantum-information processing (QIP) with this class of atomic systems~\cite{Theis+:16,Levine+:18}, 
a research direction aimed at realizing a neutral-atom quantum computer~\cite{Saffman+MoelmerRMP:10,Saffman:16}. 
In particular, the scalability milestones recently achieved with arrays of laser-cooled atoms trapped in optical microtraps 
(tweezers)~\cite{Barredo+:16,Bernien+:17,Barredo+:18,Brown+:19,deMello+:19,Schymik+:20} have rekindled interest in 
quantum-state engineering in such systems~\cite{Khazali+:16,Buchmann+:17,Ostmann+:17,Plodzien+:18,Omran+:19,Zheng+:20,
Pendse+:20,Mukherjee+:20}. Owing to the possibility of integrating multiqubit storage, readout, and transport~\cite{Beugnon+:07} 
in these systems, as well as their inherent capability for the coherent control of spin- and motional states of trapped 
atoms~\cite{Kaufman+:12}, tweezer arrays have acquired their present status of the most powerful platform for QIP with 
neutral atoms.

Rydberg blockade (RB)~\cite{RBmanybody:04,Ates+:07,RBobservation:09} -- a phenomenon whereby the van der Waals (vdW) interaction 
prevents Rydberg excitation of more than one atom within a certain radius -- established itself as the enabling physical 
mechanism for QIP with neutral atoms~\cite{Saffman+MoelmerRMP:10}. While RB also lies at the heart of many 
other phenomena~\cite{Gorshkov++:11,Dudin+Kuzmich:12}, perhaps its most important implication is that it gives rise to a 
conditional logic that enables the realization of entangling two-qubit gates~\cite{Saffman+MoelmerRMP:10,Shi:19;20},
such as controlled-NOT~\cite{Lukin+:01,Wilk+:10,Isenhower+:10}. Another important facet of RB is that it leads 
to the creation of coherent-superposition ``superatom'' states with a single Rydberg excitation 
being shared among all atoms in an ensemble~\cite{Wilk+:10}. Such entangled states~\cite{Saffman+Moelmer:09} 
belong to a special, ``twisted'' type of $W$ states~\cite{Duer+:00}. The latter represent one of the two most 
important classes of maximally-entangled multiqubit states; the other class, inequivalent with respect to local operations 
and classical communication~\cite{Nielsen:99}, is furnished by Greenberger-Horne-Zeilinger (GHZ) states~\cite{Greenberger+Horne+Zeilinger:89}.
$W$ states are known to be the most robust ones to particle losses among all $N$-qubit states~\cite{Koashi+:00} and have proven 
useful in many QIP protocols~\cite{Zhu+:15,Lipinska+:18}. This prompted proposals for their preparation in various physical 
systems~\cite{Tashima+:08,Peng+:0910,Gangat+:13,Li+Song:15,Kang+:16,StojanovicPRL:20,Sharma+Tulapurkar:20,Bugu+:20}.

While RB naturally engenders a $W$-type entanglement in ensembles of cold neutral atoms, the present work aims to show that 
the same type of entanglement can also be engineered in such systems under quite different physical circumstances. More precisely, 
this paper describes a scheme for a fast, deterministic preparation of $\pi$-twisted $W$ states in one-dimensional arrays of cold 
neutral atoms with Rydberg-dressed resonant dipole-dipole interaction~\cite{Wuester+:11}. These arrays of atoms are assumed to be 
trapped in optical tweezers~\cite{Barredo+:16,Bernien+:17,Barredo+:18,Brown+:19,deMello+:19}. 

Generally speaking, Rydberg dressing entails an off-resonant laser coupling between the ground- and Rydberg states of an atom, 
such that a small Rydberg component is admixed to the ground state~\cite{Macri+Pohl:14,Balewski+:14}. This allows the dipole-dipole 
interaction to be felt even among atoms that almost reside in their ground states. In particular, the use of Rydberg dressing in 
the realm of QIP arose from the desire to strengthen qubit-qubit interactions for Rydberg qubits encoded in two long-lived low-lying 
atomic states (typically two hyperfine sublevels of an electronic ground state~\cite{Levine+:18}). This, in turn, naturally led 
to the concept of Rydberg-dressed qubits~\cite{Petrosyan+Moelmer:14,Jau+:16,Arias+:19,Mitra+:20}.

The essential ingredient of the present work is that it explicitly takes into account the coupling of an itinerant dressed Rydberg 
excitation to the motional degrees of freedom of trapped atoms. Upon quantizing the latter into dispersionless (zero-dimensional) 
bosons, the ensuing system dynamics are governed by a nonlocal excitation-boson (e-b) interaction~\cite{Stojanovic+Vanevic:08,
Stojanovic+:14,Stojanovic+Salom:19}. In particular, for a specially chosen [``sweet-spot'' (ss)] detuning of the dressing lasers 
from the relevant internal atomic transitions, the ground state of the effective e-b Hamiltonian is the bare-excitation Bloch 
state with the quasimomentum $k=\pi$ (measured in units of the inverse period of the underlying lattice), which in the system 
at hand coincides with the $\pi$-twisted $W$ state of Rydberg-dressed qubits. 

In addition to being the ground state of the system in a broad window of the relevant parameters, the sought-after $W$ 
state is separated from the other eigenstates by a gap equal to the single-boson energy. This last circumstance -- a 
generic feature of systems in which an itinerant excitation is coupled to gapped bosons -- represents a key protection mechanism 
for the desired $W$ states and facilitates their preparation using a simple Rabi-type driving protocol. Importantly, the resulting 
state-preparation times are independent of the system size (i.e., the number of Rydberg-dressed qubits), thus the proposed system 
provides a scalable $W$-type entanglement resource. Another favorable property of the envisioned protocol is that these last times 
are several orders of magnitude shorter than the relevant Rydberg-state lifetimes.

The remainder of this paper is organized as follows. In Sec.~\ref{System} the system under consideration is introduced,
along with the necessary notation and conventions to be used throughout the paper. The effective Hamiltonian of 
the system, describing an itinerant (dressed) Rydberg excitation coupled to the motional degrees of freedom of 
atoms is discussed in Sec.~\ref{effectHamiltonian}, first in its most general form (Sec.~\ref{GenCase}) and then 
in the special case that corresponds to the ``sweet-spot'' detuning (Sec.~\ref{SweetSpot}). The principal findings of 
the paper, which pertain to this special choice of the detuning, are presented and discussed in Sec.~\ref{ResDisc}.
The character of the ground states of the system in different parameter regimes is first analyzed in Sec.~\ref{GSandWstates},
which also establishes the connection between the obtained quasimomentum-$\pi$ bare-excitation ground states  
and the desired $\pi$-twisted $W$ states. The envisioned state-preparation protocol is then presented and its robustness discussed 
in Sec.~\ref{StatePrep}. Finally, the significance of the obtained results in the context of QIP with Rydberg-dressed 
qubits is briefly elaborated on in Sec.~\ref{SignifRydbDressed}. The paper is summarized, along with conclusions and 
some general remarks, in Sec.~\ref{SummConcl}.
\section{System} \label{System}
\begin{figure}[t!]
\includegraphics[clip,width=7cm]{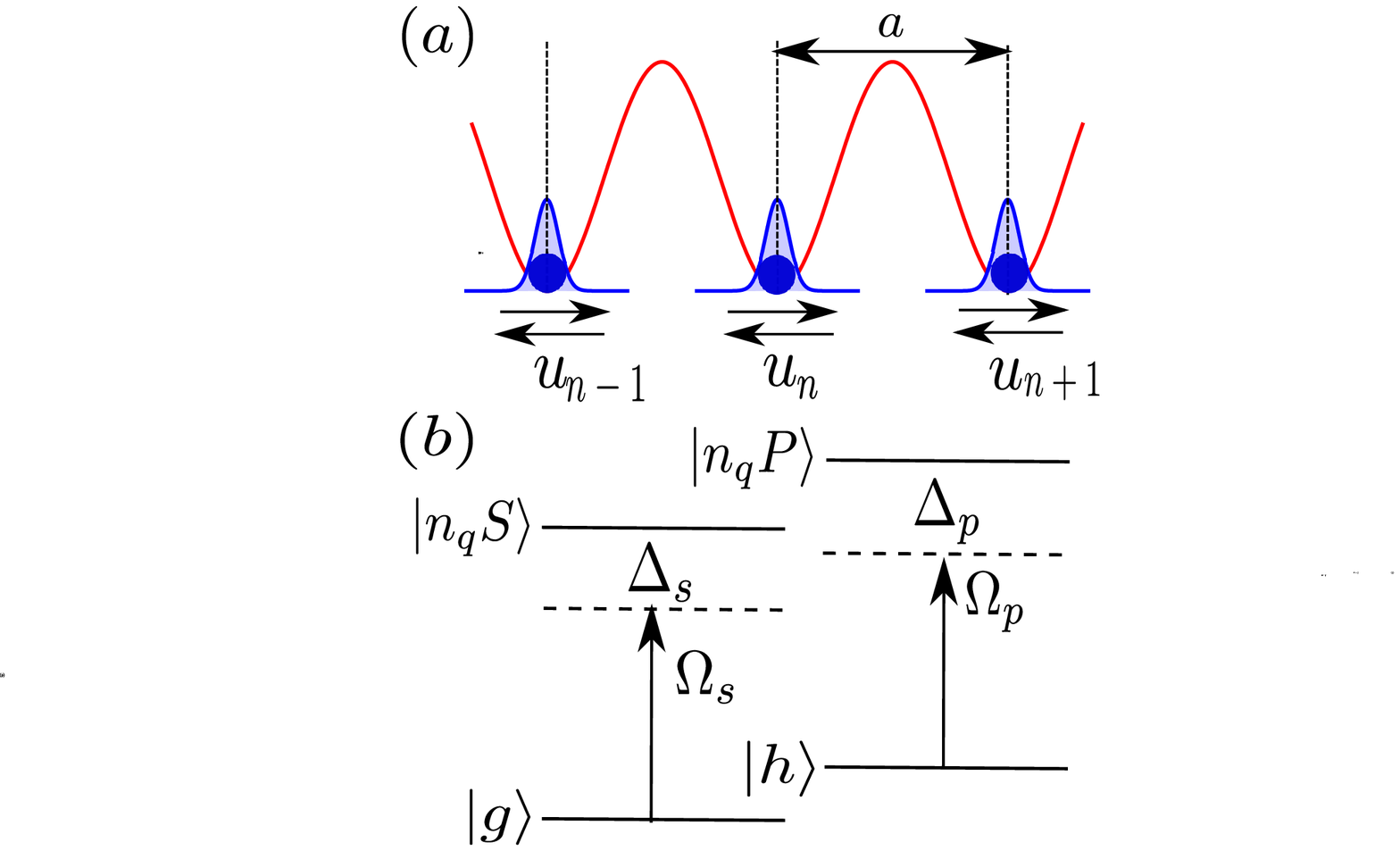}
\caption{\label{fig:system}(Color online)(a) Illustration of the system under consideration: cold neutral atoms are confined 
in individual, nearly harmonic, optical-dipole microtraps whose minima are separated by distance $a$. (b) Schematic level diagram 
of an atom with ground states $|g\rangle$ and $|h\rangle$, which are off-resonantly laser-coupled to highly-excited Rydberg states 
$|n_{\textrm q}S\rangle$ and $|n_{\textrm q} P\rangle$, respectively.}
\end{figure}

The system under consideration [for an illustration, see Fig.~\ref{fig:system}(a)] is a one-dimensional (1D) array of $N$ 
cold neutral atoms (e.g., of $^{87}$Rb) with mass $m$, each confined in its individual, approximately harmonic, optical-dipole 
microtrap. In particular, the distance $a$ between the minima of adjacent microtraps represents the period of the underlying 1D 
lattice. Importantly, the quantized displacement of the $n$-th atom ($n=1,\ldots,N$) from its equilibrium position is given 
by $u_n\equiv(\hbar/2m\omega_b)^{1/2}(b_n+b_n^{\dagger})$, where $\omega_b$ is the longitudinal trap frequency and $b_n^{\dagger}$ 
($b_n$) creates (destroys) a boson with energy $\hbar\omega_b$ in the respective microtrap. It should be stressed that an 
effectively 1D system of this kind is realized in practice by choosing the transverse trapping frequencies to be an order 
of magnitude larger than the longitudinal frequency $\omega_b$.

Unlike the vdW case, dressing resonant dipole-dipole interactions necessitates the use of two laser couplings and four electronic 
states (two ground states and two Rydberg states)~\cite{Wuester+:11}. Thus, an off-resonant coherent coupling of two ground 
states (i.e. two different levels in the hyperfine manifold of the alkali-atom electronic ground state) -- here denoted 
by $|g\rangle$ and $|h\rangle$ -- to a pair of highly excited Rydberg states $|n_{\textrm q} S\rangle$ and $|n_{\textrm q} P\rangle$ 
(where $n_{\textrm q}$ is the principal quantum number) is envisaged here [for an illustration, see Fig.~\ref{fig:system}(b)]. The 
last two states correspond to the angular-momentum quantum numbers $l=0,1$. In addition, all atoms are hereafter assumed to be 
prepared in states corresponding to the value $m_l = 0$ of the azimuthal quantum number, never acquiring $m_l \neq 0$. Along with 
the previously assumed geometric confinement of atoms, this ensures that the effective dressing-induced interaction potential for 
a pair of atoms will have no angular dependence.

As a result of Rydberg dressing, an atom that initially resided in the ground state $|g\rangle$ ($|h\rangle$) finds itself in the 
dressed state $|0\rangle\approx|g\rangle +\alpha_s|n_{\textrm q}S\rangle$ ($|1\rangle\approx |h\rangle +\alpha_p|n_{\textrm q}
P\rangle$), where $\alpha_{s,p}\equiv \Omega_{s,p}/(2\Delta_{s,p})$ are the effective (dimensionless) dressing parameters, 
fixed by the respective total Rabi frequencies $\Omega_{s,p}$ of the driving fields and the total laser detunings 
$\Delta_{s,p}$~[cf. Fig.~\ref{fig:system}(b)]~\cite{Wuester+:11}. [Note that, because the coherent coupling between the ground- 
and Rydberg states is in practice realized through two-photon- (or multi-photon-) transitions, the Rabi frequencies $\Omega_{s,p}$
and the detunings $\Delta_{s,p}$ have to be considered as effective quantities.] These parameters represent a quantitative 
measure for how far-detuned the coherent laser coupling is. For the sake of simplicity, it is hereafter assumed that 
$\alpha_s=\alpha_p\equiv\alpha$ and $\Delta_s=\Delta_p\equiv\Delta/2$, where $\Delta\equiv\Delta_s+\Delta_p$. These assumptions 
imply that $\Omega_s=\Omega_p\equiv\Omega=\Delta\alpha$.

The physical mechanism behind the coupling of the internal states of atoms in this system to their motional degrees 
of freedom represents a twofold generalization of the conventional excitation transport enabled by the resonant 
dipole-dipole interaction~\cite{Barredo+:15}; the latter scales as the inverse third power of the interatomic distance ($V_{\textrm{dd}}=C_3/R^3$).
Firstly, in the usual setting for a pair of atoms prepared in two different Rydberg states $|s\rangle$ and $|p\rangle$ 
(the latter state being higher in energy -- the ``excited state'') the resonant dipole-dipole interaction gives rise to the 
hopping of a $p$ excitation between the two atoms. Thus, this interaction is accompanied by the electronic-state transfer 
$|s,p\rangle \rightleftarrows |p,s\rangle$ between the two atoms~\cite{Ates+:08,Lienhard+:20,Abumwis+:20}. On the other hand, here the role 
of the ordinary Rydberg states is played by the dressed ones (denoted above by $|0\rangle$ and $|1\rangle$)~\cite{Wuester+:11}, 
which represent the logical qubit states. Secondly, due to the vibrational motion of atoms interatomic distances in the system 
under consideration are not fixed, but instead dynamically fluctuate, so that, e.g., the distance between atoms $n$ and $n+1$ 
is given by $a+u_{n+1}-u_n$. This leads to an effective dependence of both the excitation's on-site energy and its hopping 
amplitude on the boson degrees of freedom, two e-b coupling mechanisms reminiscent of those encountered in solid-state 
systems~\cite{Stojanovic+:04,Shneyder+:20,Hannewald++:04,Stojanovic++:10,Vukmirovic+:10}.

\section{Effective excitation-boson Hamiltonian}\label{effectHamiltonian}
In what follows, an effective e-b Hamiltonian of the system at hand is first presented (Sec.~\ref{GenCase}). The form of this Hamiltonian 
follows from the effective interaction potential for a pair of Rydberg-dressed atoms (in the resonant dipole-dipole configuration 
of their internal states), which was derived using van-Vleck perturbation theory in Ref.~\cite{Wuester+:11} (with the dresing parameters 
$\alpha_s$ and $\alpha_p$ serving as small parameters for the perturbative expansion). In the present work this last  
result is used as the point of departure for the treatment of an ensemble of atoms, with the additional assumption that the interatomic 
distances dynamically fluctuate (recall the discussion in Sec.~\ref{System}). Following the discussion of the most general e-b Hamiltonian 
of the system, further considerations are devoted to a special case of relevance for the engineering of the sought-after $W$ states 
(Sec.~\ref{SweetSpot}).
\subsection{General case}\label{GenCase}
The system Hamiltonian, describing an itinerant dressed Rydberg excitation coupled to 
dispersionless bosons, can succinctly be written as
\begin{eqnarray}\label{HamExc}
H &=&\sum_n \varepsilon_n(\mathbf{u}) c_n^{\dagger}c_n +
\sum_n t_{n,n+1}(\mathbf{u})(c^\dagger_{n+1}c_n + \mathrm{H.c.}) \nonumber\\
&+& \hbar\omega_{\textrm b}\sum_n b^\dagger_n b_n \:,
\end{eqnarray}
where $\mathbf{u}\equiv\{u_n|\:n=1,\ldots,N\}$ is a shorthand for the set of the atom 
displacements and $c^\dagger_n$ ($c_n$) creates (destroys) an excitation at site $n$, with 
\begin{eqnarray} \label{vareps}
\varepsilon_n(\mathbf{u}) &=& \frac{\alpha^4 \hbar\Delta}{2}\Bigg
[\left\{1-\left(\frac{C_3}{\hbar\Delta}\right)^2 
\frac{1}{(a+u_{n+1}-u_n)^{6}}\right\}^{-1} \nonumber \\
&+& \left\{1-\left(\frac{C_3}{\hbar\Delta}\right)^2 
\frac{1}{(a+u_{n}-u_{n-1})^{6}}\right\}^{-1}\Bigg]
\end{eqnarray}
being its corresponding on-site energy~\cite{Wuester+:11}. The latter depends on the boson displacements, 
not only on site $n$ but also on the adjacent sites $n\pm 1$. At the same time
\begin{eqnarray} \label{thopp}
t_{n,n+1}(\mathbf{u})&=&\frac{\alpha^4 C_3}{(a+u_{n+1}-u_n)^3}
\:\nonumber\\
&\times&\left\{1-\left(\frac{C_3}{\hbar\Delta}\right)^2 
\frac{1}{(a+u_{n+1}-u_n)^{6}}\right\}^{-1} 
\end{eqnarray}
is the excitation hopping amplitude between sites $n$ and $n+1$~\cite{Wuester+:11}, which depends on the difference 
$u_{n+1}-u_n$ of the respective displacements. To facilitate further analysis, it is prudent to introduce the dimensionless 
quantity $\zeta\equiv C_3/(\hbar\Delta a^3)$, the ratio of the most relevant energy scales in the system at hand. 

Before embarking on further discussion, it is useful to stress that the very existence of the dressing-induced on-site 
term in the Hamiltonian of Eq.~\ref{HamExc}, which has no analog in the standard resonant dipole-dipole interaction 
case~\cite{Ates+:08}, is a consequence of the fact that the effective dressing-induced interaction potential for a pair 
of atoms has a nonzero diagonal component~\cite{Wuester+:11}. 

For small displacements ($u_n\ll a$) it is pertinent to expand the expressions on the right-hand-side 
of Eqs.~\eqref{vareps} and \eqref{thopp} to linear order in the difference of displacements using the approximation $(1\pm x)^r 
\approx 1\pm rx$ ($|x|\ll 1$). The linear dependence of $\varepsilon_n(\mathbf{u})$ on $u_{n+1}-u_{n-1}$ captures the coupling 
of the excitation density at site $n$ with the boson displacements on the neighboring sites $n\pm 1$ [breathing-mode-type (B) e-b 
coupling]; similarly, the linear dependence of $t_{n,n+1}$ on $u_{n+1}-u_n$ describes how the excitation hopping between sites 
$n$ and $n+1$ is affected by the boson displacements [Peierls-type (P) coupling]~\cite{Stojanovic+Vanevic:08,Stojanovic+:14,Stojanovic+Salom:19}. 
This lowest-order expansion reads
\begin{eqnarray}\label{onsitehopp}
\varepsilon_n(\mathbf{u}) &=& \epsilon_0 +\xi_{\textrm{B}}(u_{n+1}-u_{n-1}) \:,\nonumber\\
t_{n,n+1}(\mathbf{u}) &=& -t_e + \xi_{\textrm P}(u_{n+1}-u_n) \:,
\end{eqnarray}
where $\xi_{\textrm{B}}$ and $\xi_{\textrm{P}}$ are given by
\begin{eqnarray}\label{xiBMP}
\xi_{\textrm{B}} &=& 3\:\frac{\alpha^4\hbar\Delta}{a}
\frac{\zeta^2}{\left(1-\zeta^2\right)^2}\:, \nonumber\\
\xi_{\textrm{P}} &=& 3\:\frac{\alpha^4 C_3}{a^4}
\frac{3\zeta^2-1}{\left(1-\zeta^2\right)^2}\:,
\end{eqnarray}
and the bare on-site energy and hopping amplitude by 
\begin{equation}\label{t_e}
\epsilon_0 = \frac{\alpha^4 \hbar\Delta}{\displaystyle 1-\zeta^2}   \:, \qquad 
t_e = -\frac{\alpha^4 C_3}{a^3\displaystyle(1-\zeta^2)} \:.
\end{equation}
The positive sign of $t_e$ [realized for $|\zeta|>1$, i.e., $C_3/(\hbar|\Delta|a^3)>1$] 
corresponds to the conventional situation where the bare-excitation dispersion $\epsilon_0-2t_e\cos k$ has 
its minimum at $k=0$. However, of principal interest here is the opposite, negative sign of $t_e$ [for $|\zeta|<1$, 
i.e., $C_3/(\hbar|\Delta|a^3)<1$]. This is the case with the band minimum at $k=\pi$, which corresponds 
to the bare-excitation Bloch state 
\begin{equation}\label{Psi_kpi}
|\Psi_{k=\pi}\rangle\equiv c^{\dagger}_{k=\pi}|0\rangle_{\textrm{e}}\otimes|0\rangle_{\textrm{b}} \:,
\end{equation}
where $|0\rangle_{\textrm{e}}$ and $|0\rangle_{\textrm{b}}$ are the excitation and boson vacuum states,
respectively.

The effective system Hamiltonian has a noninteracting part that comprises free excitation- and boson terms:
\begin{equation}\label{H_0}
H_{0}=\epsilon_0\sum_n c^\dagger_n c_n -t_e\sum_n (c^\dagger_{n+1}c_n 
+\mathrm{H.c.})+\hbar\omega_{\textrm b}\sum_n b^\dagger_n b_n \:.
\end{equation}
Its interacting part is given by $H_{\textrm{e-b}}=H_{\textrm{B}}+H_{\textrm P}$, where 
\begin{eqnarray}\label{H_BandP}
H_{\textrm{B}} &=& g_{\textrm{B}}\hbar\omega_{\textrm b}\sum_n c^\dagger_n c_n(b^\dagger_{n+1}
+ b_{n+1}- b^\dagger_{n-1}- b_{n-1})\:, \\
H_{\textrm{P}} &=& g_{\textrm P}\hbar\omega_{\textrm b}\sum_n (c^\dagger_{n+1}c_n+\mathrm{H.c.})(b^\dagger_{n+1}
+ b_{n+1}-b^\dagger_n - b_n) \nonumber\:,
\end{eqnarray}
are the terms that correspond to the two different e-b coupling mechanisms described above [cf. Eq.~\eqref{onsitehopp}],
with $g_{\textrm{B}}\equiv\xi_{\textrm{B}}/(2m\hbar\omega^{3}_{\textrm b})^{1/2}$ and 
$g_{\textrm P}\equiv \xi_{\textrm{P}}/(2m\hbar\omega^{3}_{\textrm b})^{1/2}$ being the dimensionless B- and P coupling 
strengths. Importantly, owing to the discrete translational symmetry of the system, the eigenstates of $H=H_0+H_{\textrm{e-b}}$ can 
be labelled by the eigenvalues of the total quasimomentum operator $K_{\mathrm{tot}}=\sum_{k}k\:c_{k}^{\dagger}c_{k}+
\sum_{q}q\:b_{q}^{\dagger}b_{q}$. The permissible eigenvalues of this operator belong to the Brillouin zone 
that corresponds to the underlying 1D lattice.

The bare on-site energy $\epsilon_0$ in Eq.~\eqref{H_0} plays the usual role of the on-site energy in tight-binding models 
-- that of a constant energy offset (i.e. the band-center energy) for an itinerant excitation. While $\epsilon_0$ is 
inconsequential for the physical mechanism that leads to $\pi$-twisted $W$ states in the system at hand (competition of the 
B- and P couplings), the fact that it depends on the dressing parameter $\alpha$ [cf. Eq.~\eqref{t_e}] does have some 
bearing on the preparation of such states (see Sec.~\ref{StatePrep} below).

\subsection{Special case: sweet-spot detuning}\label{SweetSpot}
Consider now the special case of the proposed system with equal P- and B coupling strengths, i.e., $g_{\textrm P}=g_{\textrm{B}}
\equiv g$. The latter physical situation corresponds to the ``sweet-spot'' (ss) value $\zeta_{\textrm{ss}}=(1+\sqrt{13})/6\approx 
0.77$ of $\zeta$. Assuming that the atomic species and the principal quantum number are chosen, which fixes the interaction constant 
$C_3$, for each choice of the lattice period $a$ this last situation is realized for the detuning $\Delta^{\textrm{ss}}\equiv 
C_3/(\hbar\zeta_{\textrm{ss}} a^3)$ and a range of values for the Rabi frequency $\Omega^{\textrm{ss}}=\Delta^{\textrm{ss}}\alpha$ 
[determined by the adopted range of values of the dressing parameter (see below)]. In this special case, the e-b-coupling and total 
Hamiltonians will be denoted by $H^{\textrm{ss}}_{\textrm{e-b}}$ and $H^{\textrm{ss}}$, respectively, in what follows. 

To quantify the e-b coupling strength in the aforementioned special case of the system under consideration, one invokes the 
momentum-space form of $H^{\textrm{ss}}_{\textrm{e-b}}$, which reads 
\begin{equation}
 H^{\textrm{ss}}_{\textrm{e-b}}=N^{-1/2}\:\sum_{k,q}\gamma^{\textrm{ss}}_{\textrm{e-b}}(k,q)\:
 c_{k+q}^{\dagger}c_{k}(b_{-q}^{\dagger}+b_{q}) \:.
\end{equation}
The explicit form of the e-b vertex function $\gamma^{\textrm{ss}}_{\textrm{e-b}}(k,q)$ in the last equation is 
\begin{equation}\label{gammass}
\gamma^{\textrm{ss}}_{\textrm{e-b}}(k,q)=2ig\hbar\omega_{\textrm b}\:[\:\sin k-\sin q-\sin(k+q)]\:.
\end{equation}
Consequently, the effective e-b coupling strength -- generally defined as $\lambda_{\textrm{e-b}}=\langle|\gamma_{\textrm{e-b}}
(k,q)|^{2}\rangle_{\textrm{BZ}}/(2|t_{\rm e}|\omega_{\textrm{b}})$~\cite{Stojanovic:20}, where $\langle\ldots\rangle_{\textrm{BZ}}$ 
stands for the Brillouin-zone average -- in this special case evaluates to $\lambda^{\textrm{ss}}_{\textrm{e-b}}
\equiv 3g^{2}\:\hbar\omega_{\textrm{b}}/|t_{\rm e}|$, i.e.,
\begin{equation}\label{lambda_param}
\lambda^{\textrm{ss}}_{\textrm{e-b}}=\frac{27}{2}\:\alpha^4\:\frac{C_3}
{m\omega_{\textrm{b}}^{2} a^5}\:\frac{(3\zeta_{\textrm{ss}}^2-1)^2}
{(1-\zeta_{\textrm{ss}}^2)^3}\:.
\end{equation}
The obtained dependence of $\lambda^{\textrm{ss}}_{\textrm{e-b}}$ on $\alpha\equiv\Omega^{\textrm{ss}}/\Delta^{\textrm{ss}}$ 
implies that the Rabi frequency $\Omega^{\textrm{ss}}$ is the main experimental knob in the system at hand. By varying  
$\Omega^{\textrm{ss}}$ different characteristic regimes of this system can be explored.

To set the stage for further analysis, it is prudent to specify at this point the realistic range of values for each of the relevant 
system parameters. The system at hand is mostly analyzed in what follows for $n_{\textrm q}=80$, with the corresponding value $C_3=2
\pi\hbar\times 40$\:GHz$\mu$m$^3$ of the resonant dipole-dipole interaction constant for $^{87}$Rb atoms. As usual for optical-tweezer 
arrays, the lattice period $a$ is in the range between about $3\:\mu$m and tens of micrometers. The corresponding values of the ss 
detuning can vary in an extremely wide range, depending on the choice of  $a$; for example, for $a=4$\:$\mu$m one obtains $\Delta^{\textrm{ss}}
\approx 5.12$\:GHz, for $a=10$\:$\mu$m one finds $\Delta^{\textrm{ss}}\approx 327.4$\:MHz, while for $a=15$\:$\mu$m one has $\Delta^{\textrm{ss}}
\approx 97$\:MHz. At the same time, the typical values for the trapping frequency $\omega_{\textrm{b}}$ are $\omega_{\textrm{b}}/(2\pi)
\sim(2 - 5)$\:kHz. Finally, for the dressing parameter $\alpha$ it is worthwhile to consider values in the range $0.01 - 0.1$.
\section{Results and Discussion} \label{ResDisc}
In the following, the principal findings of this work are presented and discussed. The character of the ground states of the system 
at hand is first analyzed (Sec.~\ref{GSandWstates}); it is explained that in a broad parameter window they coincide with the desired 
$\pi$-twisted $W$ states. The $W$-state preparation protocol is then presented, with emphasis on its robustness that stems from the 
specific character of the energy spectrum of the system (Sec.~\ref{StatePrep}). Finally, the significance of the obtained results for 
QIP with Rydberg-dressed qubits is briefly discussed in Sec.~\ref{SignifRydbDressed}.
\subsection{Ground state and its connection to $\pi$-twisted $W$ states} \label{GSandWstates}
Lanczos-type exact diagonalization~\cite{Stojanovic:20} of $H^{\textrm{ss}}=H_0+H^{\textrm{ss}}_{\textrm{e-b}}$ is carried out here for a 
system with $N=10$ sites (i.e., atoms) and the maximal number $M=8$ of bosons in the truncated boson Hilbert space. This is done using a 
well-established procedure for a controlled truncation of bosonic Hilbert spaces. This procedure entails a gradual increase of $N$, with 
the concomitant increase of $M$, until a numerical convergence of the obtained results for the ground-state energy and other relevant 
quantities is achieved~\cite{Stojanovic:20}. 

\begin{figure}[t!]
\includegraphics[clip,width=8.3cm]{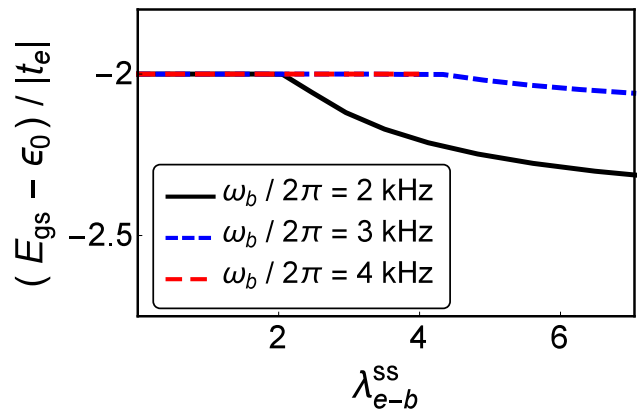}
\caption{\label{fig:Egs}(Color online)Dependence of the ground-state energy of the system, without the on-site-energy contribution 
$\epsilon_0$, on the effective coupling strength $\lambda^{\textrm{ss}}_{\textrm{e-b}}$ for $a=4$\:$\mu$m and three different 
values of the frequency $\omega_{\textrm{b}}$.}
\end{figure}

The performed numerical calculation shows that the ground state of $H^{\textrm{ss}}$ undergoes a sharp level-crossing transition~\cite{Stojanovic:20}  
at a certain critical value $(\lambda^{\textrm{ss}}_{\textrm{e-b}})_\textrm{c}$ of $\lambda^{\textrm{ss}}_{\textrm{e-b}}$.  
For $\lambda^{\textrm{ss}}_{\textrm{e-b}}<(\lambda^{\textrm{ss}}_{\textrm{e-b}})_\textrm{c}$ the ground state corresponds 
to the eigenvalue $\pi$ of $K_{\mathrm{tot}}$ and has a peculiar character. Namely, despite being the ground state of 
an interacting e-b Hamiltonian, it has the form of the bare-excitation Bloch state $|\Psi_{k=\pi}\rangle$ [cf. Eq.~\eqref{Psi_kpi}]
and its energy $\epsilon_0-2|t_{e}|$ corresponds to a minimum of a 1D tight-binding dispersion. By contrast, the strongly 
boson-dressed ground state for $\lambda^{\textrm{ss}}_{\textrm{e-b}}\ge (\lambda^{\textrm{ss}}_{\textrm{e-b}})_\textrm{c}$ is 
twofold-degenerate and corresponds to $K=\pm K_{\textrm{gs}}$, where $0<K_{\textrm{gs}}<\pi$. The dependence of the ground-state 
energy $E_{\textrm{gs}}$ (without the constant contribution $\epsilon_0$), expressed in units of $|t_{e}|$, on $\lambda^
{\textrm{ss}}_{\textrm{e-b}}$ is depicted in Fig.~\ref{fig:Egs}.

\begin{figure}[b!]
\includegraphics[clip,width=8.3cm]{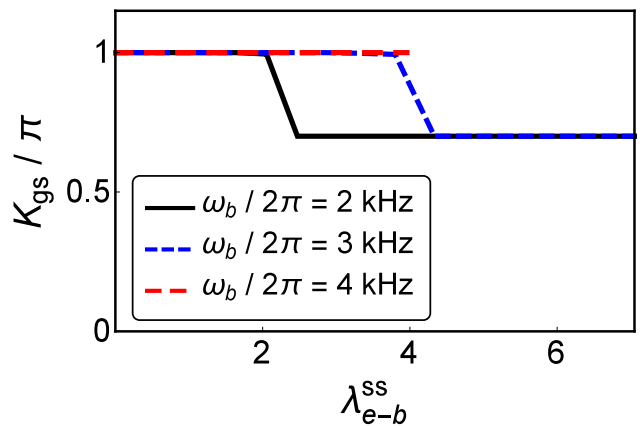}
\caption{\label{fig:Kgs}(Color online)Dependence of the ground-state total quasimomentum on the effective coupling 
strength $\lambda^{\textrm{ss}}_{\textrm{e-b}}$ for $a=4$\:$\mu$m and three different values of the frequency 
$\omega_{\textrm{b}}$.}
\end{figure}

Figure~\ref{fig:Kgs} illustrates how the ground-state total quasimomentum $K_{\textrm{gs}}$ depends on $\lambda^{\textrm{ss}}_{\textrm{e-b}}$. 
In particular, $K_{\textrm{gs}}=\pi$ in the bare-excitation ground state $|\Psi_{k=\pi}\rangle$ has a vanishing bosonic 
contribution as $\langle\Psi_{k=\pi}|\sum_n b^\dagger_n b_n|\Psi_{k=\pi}\rangle=0$. The fact that this bare-excitation 
Bloch state is a ground-state of an interacting e-b system is a direct implication of the assumption that $g_{\textrm P}=
g_{\textrm{B}}\equiv g$ (recall Sec.~\ref{SweetSpot}), i.e. it is a consequence of an effective mutual cancellation of 
P- and B couplings for a bare excitation with this particular quasimomentum ($k=\pi$).

It is pertinent at this point to establish  the connection between the ground states of the system at hand and the 
desired $N$-qubit $W$ states. The bare-excitation Bloch state $|\Psi_{k}\rangle$, recast in terms of the pseudospin-$1/2$ 
(qubit) degrees of freedom, coincides with the twisted $W$ state 
\begin{equation}\label{twistedW}
|W_N(k)\rangle =N^{-1/2}\sum_{n=1}^{N}\:e^{ikn}|0\ldots 1_n\ldots 0\rangle
\end{equation}
parameterized by the quasimomentum $k$ from the Brillouin zone (i.e. $-\pi< k \leq \pi$). In particular, $|\Psi_{k=\pi}\rangle$ -- the 
ground state of the system at hand for $\lambda^{\textrm{ss}}_{\textrm{e-b}}<(\lambda^{\textrm{ss}}_{\textrm{e-b}})_
\textrm{c}$ -- corresponds to the $\pi$-twisted $W$ state $|W_N(k=\pi)\rangle$ of Rydberg-dressed qubits. 

The conditions for realizing the desired states $|W_N(k=\pi)\rangle$ in the system under consideration are easily 
reached with realistic values of the relevant experimental parameters ($a,\omega_{\textrm{b}},\alpha$). To justify 
that, it is worthwhile to immediately note that Figs.~\ref{fig:Egs} and \ref{fig:Kgs} correspond to $a=4\:\mu$m, a 
relatively small lattice period which favors larger coupling strengths [cf. Eq.~\eqref{lambda_param}] and allows the 
onset of a sharp transition. Yet, already for this choice of $a$, with a sufficiently large trapping frequency 
($\omega_{\textrm{b}}\gtrsim 2\pi\times 3.5$\:KHz) the effective coupling strength $\lambda^{\textrm{ss}}_{\textrm{e-b}}$ 
is always below the critical one, i.e., $\pi$-twisted $W$ states are accessible in the entire adopted range of values 
($0.01-0.1$) for the dressing parameter $\alpha$. The fast decay of $\lambda^{\textrm{ss}}_{\textrm{e-b}}$ with $a$ ensures 
that for $a\gtrsim 5\:\mu$m the sought-after $W$ states are the ground states of the system for any realistic choice of 
$\omega_{\textrm{b}}$ and $\alpha$. For the sake of completeness, it is worthwhile mentioning that by choosing a smaller
principal quantum number these conditions are even easier to satisfy because of the smaller value of $C_3$; for instance, 
for $n_{\textrm q}=50$ the corresponding value of this interaction constant for $^{87}$Rb is an order of magnitude 
smaller than for $n_{\textrm q}=80$.

\begin{figure}[t!]
\includegraphics[clip,width=7.6cm]{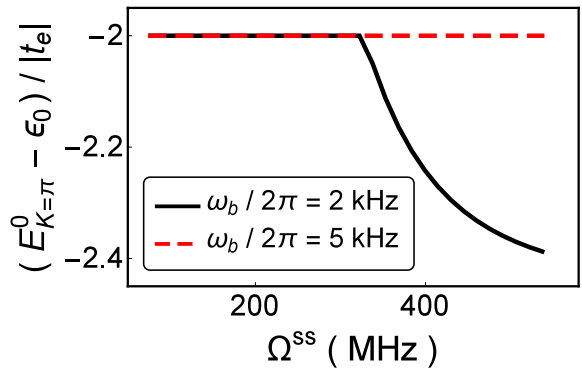}
\caption{\label{fig:E_pi}(Color online)Dependence of the lowest-energy $K=\pi$ eigenvalue (without the on-site-energy 
contribution $\epsilon_0$), whose flat part at the energy $-2|t_{\textrm{e}}|$ corresponds to the $\pi$-twisted 
$W$ state, on the Rabi frequency $\Omega^{\textrm{ss}}$ for $a=4$\:$\mu$m and two different trapping frequencies.}
\end{figure}

It is interesting to observe that -- in addition to being the ground state of $H^{\textrm{ss}}$ for coupling strengths 
below the critical one -- the state $|\Psi_{k=\pi}\rangle$ is an exact eigenstate of this Hamiltonian for an arbitrary 
$\lambda^{\textrm{ss}}_{\textrm{e-b}}$. Namely, given that $\gamma^{\textrm{ss}}_{\textrm{e-b}}(k=\pi,q)=0$ for an arbitrary 
$q$ [cf. Eq.~\ref{gammass}], it is straightforward to show that $H^{\textrm{ss}}_{\textrm{e-b}}|\Psi_{k=\pi}\rangle=0$. 
Thus, $|\Psi_{k=\pi}\rangle$ is an eigenstate of $H^{\textrm{ss}}_{\textrm{e-b}}$. Because this last state is an eigenstate 
of the free Hamiltonian $H_0$ as well, it follows immediately that it is also an eigenstate of the total Hamiltonian 
$H^{\textrm{ss}}=H_0+H^{\textrm{ss}}_{\textrm{e-b}}$. To conclude, even for those parameters (i.e., values of the Rabi frequency 
that lead to coupling strengths above the critical one) for which $|\Psi_{k=\pi}\rangle$ does not coincide with the 
lowest-energy $K=\pi$ eigenstate of the system (for an illustration, see Fig.~\ref{fig:E_pi}) this state still remains an 
eigenstate in the discrete (bound-state) part of the spectrum of $H^{\textrm{ss}}$.

\subsection{$W$-state preparation protocol} \label{StatePrep}
A deterministic Rabi-type driving protocol for the preparation of $\pi$-twisted $W$ state is discussed in what follows, 
assuming that the initial state of the system is $|0\rangle_{\textrm{e-b}}\equiv|0\rangle_{\textrm{e}}\otimes|0\rangle_{\textrm{b}}$, 
where $|0\rangle_{\textrm{e}}$ is the shorthand for an $N$-atom state with zero Rydberg-dressed excitations (i.e., all atoms 
occupying the dressed state $|0\rangle$), while $|0\rangle_{\textrm{b}}$ is the collective boson vacuum, i.e. the state with all 
atoms being in their motional ground state. For the system to be prepared in the state $|0\rangle_{\textrm{e-b}}$, two preliminary 
steps ought to be carried out. The first one is to prepare the state $|0\rangle_{\textrm{e}}$ via Rydberg-dressing starting from 
the state $|g\rangle_{\textrm{e}}$ with all atoms in their absolute ground state $|g\rangle$. The other one is to prepare all atoms 
in their motional ground states using the established methodology for this purpose~\cite{Kaufman+:12}. 

The envisioned state-preparation protocol is based on an external driving given by
\begin{equation}\label{driveop}
F_{q_d}(t)=\frac{\hbar\beta(t)}{\sqrt{N}}\sum_{n=1}^{N}\left
(\sigma_{n}^{+}e^{-iq_d n}+\sigma_{n}^{-}e^{iq_d n}\right) \:,
\end{equation}
where $\beta(t)$ accounts for its time dependence and the factors $e^{\pm iq_d n}$ allow for the possibility 
of applying external driving to different Rydberg-atom qubits with a nontrivial phase difference. The transition 
matrix element of $F_{q_d}(t)$ between $|0\rangle_{\textrm{e-b}}$ and $|\Psi_{k=\pi}\rangle\equiv |W_N(k=\pi)
\rangle\otimes|0\rangle_{\textrm{b}}$ is equal to $\hbar\beta(t)\:\delta_{q_d,k=\pi}$, indicating that the required phase 
difference between adjacent qubits is $q_d=\pi$. Furthermore, by assuming that $\beta(t)=2\beta_{p}\cos(\omega_{d}t)$, where 
$\hbar\omega_{d}\equiv\epsilon_0-2|t_{e}|$ is the energy difference between the two relevant states, in the rotating-wave 
approximation these states are Rabi-coupled with the Rabi frequency $\beta_{p}$~\cite{ShoreBookVol1}. Therefore, $\pi$-twisted 
$W$ states are prepared within time $\tau_{\textrm{prep}}=\pi\hbar/(2\beta_{p})$, which does not depend on the system size ($N$) 
at all. For example, taking $\beta_{p}/(2\pi\hbar)$ to be in the range $(10 - 100)$\:MHz, one finds $\tau_{\textrm{prep}}\approx 3 
- 25$\:ns, which is $4 - 5$ orders of magnitude shorter than typical lifetimes of ordinary Rydberg states and $7 - 8$ orders 
than those of Rydberg-dressed states, as they are scaled by an additional factor of $\alpha^{-2}$.

Apart from being the ground state of the system in a broad parameter window and remaining its eigenstate even 
outside of that window, the desired $W$ state has another property that facilitates its preparation. Namely, it is separated 
from the other eigenstates of $H^{\textrm{ss}}$ by an energy gap of $\hbar\omega_{b}$. Systems in which dispersionless bosons 
interact with a single itinerant particle~\cite{Stojanovic+:14,Stojanovic+Salom:19} generically posses such a gap, equal to 
the single-boson energy, which separates their ground state from the one-boson continuum (inelastic scattering threshold). In 
the weak-coupling regime ground states of such systems are typically the only bound states they have~\cite{Stojanovic+:14}. 
Importantly, apart from increasing the parameter window where $W$ states can be engineered, another advantage of increasing 
$a$ is a better separation of those states from other states. Namely, an increase of $a$ leads to a decrease of $\omega_{d}
\propto\alpha^4 C_3/a^3$, so that $\hbar\omega_{b}$ becomes a progressively larger fraction of the energy difference 
$\hbar\omega_{d}$. For instance, with $\alpha=0.05$ and $a=15\:\mu$m, for the chosen range of trapping frequencies $\omega_{b}$
the gap energy amounts to $15 - 40\:\%$ of this energy difference, which ensures that the above Rabi-type state-preparation 
protocol will not be hampered by an inadvertent population of undesired states. 

The analysis of the ground-state properties of the system under consideration in Sec.~\ref{GSandWstates} mostly featured the 
results that correspond to the relatively small latice period $a=4$\:$\mu$m. It is important to stress that this choice, which favors 
large effective e-b coupling strengths and a possible onset of a sharp transition [cf. Sec.~\ref{GSandWstates}], was intentionally made
in order to highlight the worst-case scenario as far as the realization of the desired $W$-type entanglement resurce in the system at 
hand is concerned. However, from the point of view of an actual $W$-state preparation, for the reasons stated above it is more favorable
to choose an intermediate or large lattice period, say $a\gtrsim 12\:\mu$m. Not only that this precludes an inadvertent population 
of undesired states in the continuum part of the spectrum of the relevant coupled e-b system, but it also leads to smaller values 
of the ss detuning (note that $\Delta^{\textrm{ss}}\propto a^{-3}$), which makes this last detuning far smaller than the typical
energy spacings of Rydberg levels. [Recall that the distribution of Rydberg energy levels becomes denser as the principal quantum 
number $n_q$ increases, such that the energy difference $\Delta E$ of adjacent levels scales as $n_q^{-3}$; note also that $\Delta E 
\sim 1$\:GHz for $n_q\sim 100$.~\cite{GallagherBOOK}] This, in turn, prevents the possibility of inadvertently populating higher 
Rydberg levels in the initial, Rydberg-dressing step (i.e. in the preliminary preparation of the state $|0\rangle_{\textrm{e}}$) of 
the proposed $W$-state preparation protocol.
\subsection{Significance for quantum-information processing with Rydberg-dressed qubits} \label{SignifRydbDressed}
Owing to the rich energy-level structure of Rydberg atoms and the existing wealth of techniques for the 
coherent manipulation of atomic internal states, there are several possibilities for storing and manipulating 
quantum information, i.e., QIP with Rydberg qubits. Depending on the number of ground- or Rydberg 
states that make up the qubit (i.e., serve as its logical $|0\rangle$ and $|1\rangle$ states), there 
are three main types of Rydberg qubits: ($\imath$) those based on one weakly-interacting state $|g\rangle\equiv|0\rangle$
and one strongly-interacting Rydberg state $|r\rangle\equiv|1\rangle$ ($gr$-qubits), ($\imath\imath$) 
those encoded using two different Rydberg states ($rr$-qubits, where $|r\rangle\equiv|0\rangle$ and
$|r'\rangle\equiv|1\rangle$), and, finally, ($\imath\imath\imath$) those encoded in two long-lived 
low-lying atomic states $|g\rangle\equiv|0\rangle$ and $|h\rangle\equiv|1\rangle$ ($gg$-qubits). 

In particular, $gg$-qubits typically involve two (usually magnetically insensitive) hyperfine sublevels of the 
electronic ground state -- like the states $|g\rangle$ and $|h\rangle$ of the system under consideration [cf. 
Fig.~\ref{fig:system}(b)]. Such qubits offer the best trade-off between coherence times and switchable interactions, 
which makes them promising candidates for universal quantum computing. On the other hand, compared to their $gr$- 
and $rr$ counterparts, $gg$-qubits are weakly interacting. One possible approach for inducing stronger interactions 
between such qubits relies on momentarily exciting and de-exciting them via Rydberg states using precisely timed 
or shaped optical fields. An alternative approach for making $gg$-qubits interact more strongly -- of relevance 
for the present work -- is to weakly admix some Rydberg-state character to the ground states using an off-resonant 
laser coupling, thereby effectively transforming them into Rydberg-dressed qubits~\cite{Petrosyan+Moelmer:14,Jau+:16,Arias+:19,Mitra+:20}. 

Generally speaking, creating quantum entanglement in large systems on timescales much shorter than the relevant coherence 
times is key to efficient QIP. In particular, it is shown here that in neutral-atom arrays $\pi$-twisted $W$ states 
of Rydberg-dressed qubits can be engineered with the corresponding preparation times being independent of the system size. 
Importantly, those preparation times are several orders of magnitude shorter than the typical lifetimes of the relevant 
Rydberg states, being at the same time an even much smaller fraction of the effective lifetimes of their Rydberg-dressed 
counterparts.

Another favorable feature of the system at hand -- and a prerequisite for universal quantum computation -- stems from 
the $XY$ character~\cite{Schuch+Siewert:03} of the effective qubit-qubit interaction in this system, which is given by 
the free-excitation hopping term in Eq.~\eqref{H_0}. and the Peierls-coupling term in Eq.~\eqref{H_BandP}. When recast in terms of 
the pseudospin-$1/2$ degrees of freedom of Rydberg-dressed qubits, the coupling between qubits $n$ and $n+1$ is given by $J_{n,n+1}
(\sigma^{x}_n \sigma^{x}_{n+1}+\sigma^{y}_n\sigma^{y}_{n+1})$, where $J_{n,n+1}\equiv 2[-t_e+g_{\textrm P}\hbar\omega_{\textrm b}
(b^\dagger_{n+1}+ b_{n+1}-b^\dagger_n - b_n)]$ is the effective $XY$-coupling strength that depends dynamically on the boson 
degrees of freedom. Such boson-dependent coupling strengths are characteristic, for instance, of certain trapped-ion systems 
where collective motional modes (phonons)~\cite{TrappedIonPRLs} play the role of bosons~\cite{Wall+:17}. Unlike such trapped-ion 
systems, whose phonon spectra have a quasicontinuous character~\cite{TrappedIonPRLs}, the system at hand merely involves 
dispersionless bosons of one single frequency. This circumvents the spectral-crowding problem that poses an obstacle for QIP 
in large trapped-ion chains~\cite{Landsman+:19}. 
\section{Summary and Conclusions} \label{SummConcl}
This work proposed a scheme for a deterministic creation of a large-scale $W$-type 
entanglement in optical tweezer arrays of atoms with Rydberg-dressed resonant dipole-dipole interaction. The resulting $W$-state 
preparation times are independent of the system size, being also several orders of magnitude shorter than the effective lifetimes 
of the relevant atomic states. It is demonstrated here that the mechanism behind this scalable entanglement resource is robust 
against the unavoidable coupling of an itinerant dressed Rydberg excitation with the motional degrees of freedom of atoms. Another 
argument in favor of the robustness of the proposed scheme stems from the fact that $\pi$-twisted $W$ states that it aims to realize 
represent ground states of the underlying system, which are at the same time separated from their other eigenstates by a sizeable 
spectral gap. The recent advances in the manipulation, control, and readout of neutral-atom states in optical dipole traps~\cite{Shea+:20} 
and the scalability of tweezer arrays bode well for an experimental implementation of this scheme.

In atomic physics, motional degrees of freedom~\cite{Morinaga+:99,Buchkremer+:00} have long been perceived exclusively 
as sources of decoherence and dephasing~\cite{Li++:13}, and have only in recent years been viewed as a quantum 
resource~\cite{Buchmann+:17}. The present work constitutes a demonstration as to how the influence of motional degrees of 
freedom can be suppressed for the sake of carrying out specific QIP tasks, even in systems where they may play a useful role 
in other tasks. This particular aspect of the present work could be generalized to other systems, such as trapped Rydberg 
ions~\cite{Higgins+:17}. While the latter have quite recently attracted considerable attention in the context of time-efficient 
gate realizations~\cite{Zhang+:20}, quantum-state engineering in such systems is a largely unexplored subject.

The concept of Rydberg dressing has in recent years been utilized in diverse contexts~\cite{Jau+:16}. In particular,
the present work underscores its usefulness in engineering maximally-entangled states of Rydberg-dressed qubits. In this sense 
the present study is complementary to that of Ref.~\cite{Buchmann+:17}, which discussed the preparation of various motional 
states of Rydberg-dressed atoms. Along with their known advantage -- namely, that their effective lifetimes 
are significantly longer than those of ordinary Rydberg states -- the capability of creating entanglement in an ensemble of atoms 
provides additional motivation to consider QIP with Rydberg-dressed states. Experimental realizations of the proposed $W$-state 
preparation scheme -- as well as theoretical explorations of its possible generalizations -- are clearly called for.
\begin{acknowledgments}
This research was supported by the Deutsche Forschungsgemeinschaft (DFG) -- SFB 1119 -- 236615297.
\end{acknowledgments}


\end{document}